\documentclass[acmtog]{acmart}

\usepackage{graphicx}
\usepackage{bm}
\usepackage{colortbl}
\usepackage{amsmath}
\DeclareMathOperator*{\argmin}{arg\,min}

\copyrightyear{2026}
\acmYear{2026}
\setcopyright{cc}
\setcctype{by}
\acmConference[SIGGRAPH Conference Papers '26]{Special Interest Group on Computer Graphics and Interactive Techniques Conference Conference Papers}{July 19--23, 2026}{Los Angeles, CA, USA}
\acmBooktitle{Special Interest Group on Computer Graphics and Interactive Techniques Conference Conference Papers (SIGGRAPH Conference Papers '26), July 19--23, 2026, Los Angeles, CA, USA}
\acmDOI{10.1145/3799902.3811075}
\acmISBN{979-8-4007-2554-8/2026/07}

\setcitestyle{square}

\newcommand{\change}[1]{#1}

\begin{document}

\title{Shellular Metamaterial Design via Compact Electric Potential Parametrization}

\author{Tianyi Huang}
\email{twisland@outlook.com}
\orcid{0009-0007-9773-1174}

\author{Chang Liu}
\email{willcliu@nus.edu.sg}
\orcid{0000-0001-5561-7137}

\author{Bohan Wang}
\email{bh.wang@nus.edu.sg}
\orcid{0000-0003-1439-1455}
\affiliation{%
  \institution{National University of Singapore}
  \country{Singapore}
}

\renewcommand{\shortauthors}{Huang et al.}

\begin{abstract}

We present a compact yet highly expressive design space for shellular metamaterials 
that support both interactive exploration and inverse design.
With only a few dozen charges, 
our representation generates a wide family of periodic shells, 
spanning from simple planar configurations to complex TPMS-like morphologies.
To enable rapid evaluation, 
we introduce an efficient GPU-based homogenization pipeline 
that computes the effective elastic tensor of a candidate design in near real time ($\sim0.4\,\mathrm{s}$), 
making interactive shellular design practical.
Across a large set of synthesized structures, 
our design space exhibits geometric diversity and 
spans a broad spectrum of mechanical responses, 
covering a wide range of effective material properties.
This fast evaluation further enables inverse design for target macroscopic properties.
In the low-solid-volume regime, 
the resulting shellular structures achieve performance competitive with state-of-the-art shell-based metamaterials in multiple material properties.
Finally, we validate manufacturability by fabricating tiled prototypes via additive manufacturing, demonstrating the
potential of our approach for real-world engineering applications.

\end{abstract}


\begin{teaserfigure}
\centering
\includegraphics[width=0.9\textwidth]{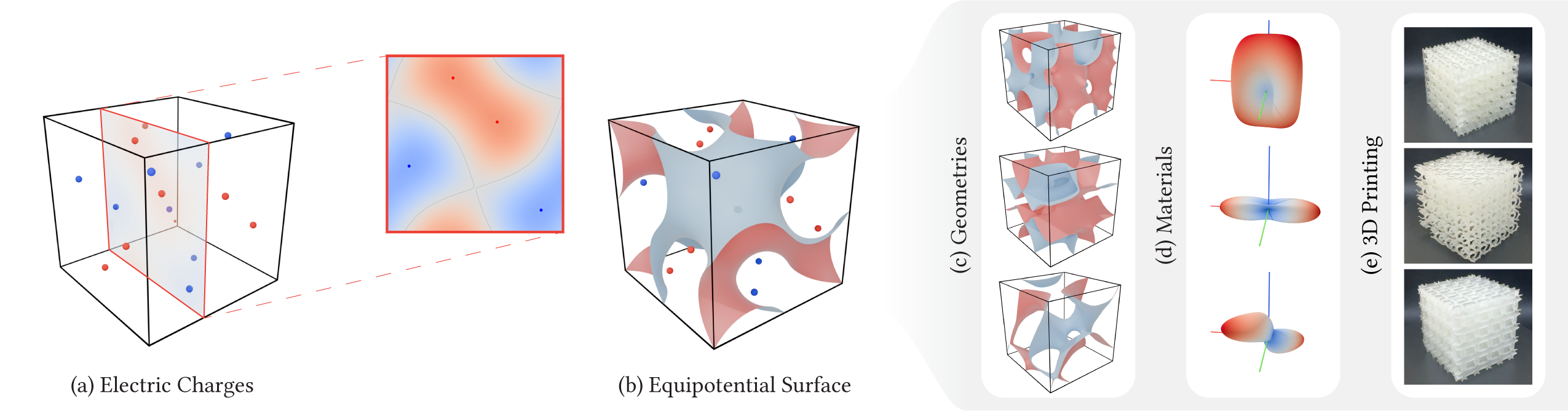}
\vspace{-0.1cm}
\caption{\textbf{Design of shellular metamaterials.} Our structures are represented by an implicit function parameterized by a few dozen electric charges (a). 
By arranging these charges, we obtain a broad set of geometrically diverse structures (b,c). 
These structures also exhibit a wide range of properties (d) and remain feasible for fabrication in practical applications (e).}
\label{fig:teaser}
\end{teaserfigure}


\maketitle

\section{Introduction}

Metamaterials have long been of interest because they exhibit material properties that differ from those of their constituent materials. 
Rather than relying solely on chemical composition, 
these properties arise predominantly from the structural arrangement of cellular architectures~\cite{babaee20133d, fang2018programmable, mao2020designing,Tang_2023}.
Within the family of 3D metamaterials, 
shellular structures stand out due to their minimal volumetric occupation and advantageous geometric features. 
These characteristics enable significant material savings while preserving structural integrity. 
In some instances, shellular structures can even outperform other architecture types~\cite{torquato2002multifunctional}. 
As a result, shellular structure designs have emerged as key elements in numerous applications, 
including 3D-printed infill patterns, heat exchanger interfaces, and reconfigurable thin-shells~\cite{huang2024shock, tozoni2024cut}.

Despite substantial progress in metamaterial design, 
designing \emph{shellular} metamaterials remains challenging.
Many automated approaches rely on restrictive design spaces that are not suitable for thin-shell-like microstructures.
For example, Panetta et al.~\shortcite{panetta2015elastic} focus on truss-based architectures, 
leaving surface-based shellular structures largely unexplored.
Methods that do target shellular structures often require solving intricate optimization or simulation problems within a narrow family of templates, 
resulting in high computational cost and limited property diversity~\cite{xu2023design,Panetta_2017,hsieh2019mechanical,Zhang_2023,Montes_Maestre_2024,Li_2023,Tozoni_2020,Xue_2025}.
At the other extreme, procedural and voxel-based representations aim for broad geometric flexibility, 
but each comes with drawbacks: 
procedural generators typically do not support both efficient structure synthesis and automatic inverse
design~\cite{makatura2023procedural},
while voxel-based representations introduce an excessively large number of degrees of freedom (DOF)~\cite{eschenauer2001topology,Liu_2018},
inheriting the scalability and regularization challenges of conventional topology optimization.

To address these limitations, 
we propose a novel design space that is both computationally efficient and sufficiently expressive to capture a broad range of shellular structures.
Our approach simultaneously achieves geometric diversity and tunable material properties, 
demonstrating great potential for shellular metamaterial design.
Additionally, we introduce a highly efficient, GPU-based computation and optimization framework,
enabling rapid structure evaluation.
Furthermore, we develop a design and optimization method that, given user-defined objectives, can automatically identify the desired structure.
Collectively, these contributions significantly expand the design possibilities for shellular metamaterials beyond current constraints.

Finally, we validate the proposed approach through comprehensive experiments.
First, we show that by randomly generating various shellular structures, 
our representation spans a broad range of material properties,
including Young's modulus, bulk modulus, shear modulus, and anisotropy of the elastic tensor.
Our evaluations are performed efficiently, 
requiring on average less than 0.4\,s to generate both a structure and its homogenized elastic tensor with high accuracy,
enabling interactive applications.
Moreover, our representation accommodates geometries ranging from simple planar shapes to more intricate shellular forms,
such as Triply Periodic Minimal Surfaces (TPMS).
We also illustrate the capability of our framework to design materials according to specific objectives,
including maximizing bulk modulus, maximizing shear modulus, or achieving a targeted Young's modulus.
\change{Several of our designs reaching up to 99.8\% in uniaxial stiffness, 
corresponding to a plane-like configuration. 
Finally, we fabricate multiple examples to demonstrate potential manufacturability.}
\vspace{-0.2cm}
\section{Related Work}
\subsection{Shellular Metamaterials}
In this paper, we focus on \emph{shellular} metamaterials, 
a particular type of mechanical metamaterial that usually fits within a \emph{cubic} unit cell 
and is composed of a \emph{homogeneous} elastic material forming thin shells. 
These shells can constitute either a two-dimensional manifold surface embedded in three-dimensional space~\cite{nguyen2016optimal} 
or a non-manifold or non-smooth surface~\cite{spadoni2014closed}. 
A prominent example within this class is the family of TPMS, 
which have been extensively studied and deployed in various applications~\cite{han2018overview,Zhang_2025},
because of its excellent mechanical and additive-manufacturing properties.
Beyond ``true'' TPMS, 
there are TPMS-like structures that share similar qualities~\cite{hu2020efficient,xu2023new,jiang2023meshless}.
Our proposed design space is capable of incorporating a broad range of these TPMS structures.

\begin{table}
    \centering
    \resizebox{\columnwidth}{!}{
    \begin{tabular}{l|cccc}
    \hline
        Representation & DOFs & Periodicity & Solving & Intuitiveness \\
        \hline
        Voxel-based       & \cellcolor{red!15}   High & \cellcolor{red!15}   No  & \cellcolor{green!15} No  & \cellcolor{red!15}   No  \\
        Fourier basis     & \cellcolor{green!15} Low  & \cellcolor{green!15} Yes & \cellcolor{green!15} No  & \cellcolor{red!15}   No  \\
        RBF basis         & \cellcolor{green!15} Low  & \cellcolor{red!15}   No  & \cellcolor{red!15}   Yes & \cellcolor{green!15} Yes \\
        PDE (field-based) & \cellcolor{green!15} Low  & \cellcolor{red!15}   No  & \cellcolor{red!15}   Yes & \cellcolor{green!15} Yes \\
        PDE (mesh-based)  & \cellcolor{green!15} Low  & \cellcolor{red!15}   No  & \cellcolor{red!15}   Yes & \cellcolor{green!15} Yes \\ 
        Boundary loop     & \cellcolor{green!15} Low  & \cellcolor{red!15}   No  & \cellcolor{red!15}   Yes & \cellcolor{green!15} Yes \\
        Our method        & \cellcolor{green!15} Low  & \cellcolor{green!15} Yes & \cellcolor{green!15} No  & \cellcolor{green!15} Yes \\
        \hline        
    \end{tabular}
    }
    \caption{
    \textbf{Comparison of design-space types for shellular structures.}
    Green denotes a desirable attribute (low DOFs, periodicity by construction, no numerical solves, and intuitive user control),
    while red denotes the opposite.
    Column \emph{DOFs} indicates the number of design DOFs.
    \emph{Periodicity} indicates whether periodic boundary conditions are enforced by construction.
    \emph{Solving} indicates whether numerical solves (e.g., PDE or linear system solves) are required during synthesis or evaluation.
    \emph{Intuitiveness} indicates whether users can intuitively and directly steer the resulting designs.
    }
    \vspace{-1cm}
    \label{tab:design-space}
\end{table}
\vspace{-0.2cm}
\subsection{Design Space}
Table~\ref{tab:design-space} summarizes several commonly used design-space representations for shellular microstructures and
highlights their trade-offs.
\vspace{-0.4cm}
\paragraph{Voxel-based representations.}
A straightforward approach is to represent geometry on an Eulerian grid (e.g., a sampled signed distance field)~\cite{osher2003constructing}.
This representation is highly general, 
but the voxel values themselves become the design variables, 
so the number of DOFs grows cubically with resolution. 
As a result, representing thin shells without severe discretization artifacts typically requires very high resolutions, 
making the design space prohibitively large. 
Learning-based compression and model-reduction methods can reduce the effective dimensionality~\cite{park2019deepsdf},
but they usually rely on curated training datasets of shellular metamaterials, which can be expensive to obtain.
\vspace{-0.2cm}
\paragraph{Basis expansions.}
To obtain a compact design space, one can parameterize an implicit field using a small set of basis functions.
A common example is a truncated Fourier expansion that retains only low-frequency terms (``Fourier basis'' in Table~\ref{tab:design-space}).
This choice enforces periodicity by construction, 
but directly manipulating Fourier coefficients (``intuitiveness'' in Table~\ref{tab:design-space}) is often not straightforward:
enforcing symmetries typically requires carefully tying multiple coefficients, 
and capturing localized geometric features can require many modes.
In contrast, our representation provides a structured low-dimensional parameterization of a periodic implicit field, 
where a small number of charge parameters jointly control many coupled modes, 
enabling intuitive geometric steering while retaining compactness.
In our experiments, we further observe that, 
for a comparable number of design DOFs, 
our parameterization spans a broader variety of shellular surfaces than a truncated Fourier representation (Sec.~\ref{sec:structure-gen}).
One could likewise build an implicit field from the analytic periodic Green’s function of Laplace’s equation with point
heat sources; 
this offers similarly interpretable “handle”-based control, 
but when used as a basis (e.g., with only source positions as variables) 
it provides less diversity in material properties than our formulation,
as shown in Sec.~\ref{sec:structure-gen}.

Radial basis function (RBF) representations similarly offer compact and intuitive control by 
placing a small number of handles (centers) in Euclidean space and adjusting their weights. 
However, periodicity is not inherent and typically requires additional machinery (e.g., tiling
centers or using periodic kernels). 
Moreover, in many RBF constructions the coefficients are obtained by solving an auxiliary fitting/interpolation problem, 
which introduces additional computational cost when the handles are updated.
\vspace{-0.2cm}
\paragraph{PDE-based representations.}
Another line of work uses PDEs either to deform a given surface template (``PDE (mesh-based)'') or 
to define an implicit field whose isosurface represents geometry (``PDE (field-based)''). 
Mesh-based approaches build deformation bases from geometric energies (e.g., biharmonic energy~\cite{wang2015linear}),
parametric surface models (e.g., NURBS~\cite{piegl2012nurbs}), or related shape-modeling tools, 
providing intuitive handles but requiring a predefined mesh template with fixed topology.
Moreover, incorporating periodic constraints is also nontrivial.
Field-based approaches use boundary conditions as control variables and extract geometry from the resulting PDE solution, 
for instance by using heat sources and Laplace's equation~\cite{ulu2019structural,weizheng2025dualms}. 
While these methods can provide intuitive control, 
they generally require numerical solves as parameters change, 
limiting interactive exploration and large-scale inverse design.
Phase-field approaches can also define shell-like structures, 
e.g., via the Cahn-Hilliard equation~\cite{hsieh2019mechanical}, 
but the required PDE solves can be computationally expensive~\cite{kumar2020inverse}.
\vspace{-0.2cm}
\paragraph{Boundary-loop representations.}
An alternative is to specify thin shells via boundary loops and then construct surface patches that interpolate these curves~\cite{makatura2023procedural,xu2023new,palmer2022deepcurrents}.
However, generating patches from loops typically involves solving multiple optimization problems (often taking seconds to minutes), 
which makes large-scale data acquisition cumbersome. 
In addition, authoring and parameterizing loops is itself nontrivial; 
to simplify the problem, many approaches define loops within a fundamental domain and replicate patches across the unit cell, 
which can restrict the achievable symmetries and structural diversity.
\vspace{-0.2cm}
\paragraph{Voronoi-based parameterizations.}
\change{
A recent approach generates shellular structures from a generalized Voronoi partition defined by a set of sites and anisotropic distance metrics~\cite{Numerow_2024,numerow2025star}. 
This formulation provides a differentiable shell design framework for targets such as nonlinear stress-strain behavior and auxetic responses.
However, the resulting structures are inherently piecewise planar, leading to nonsmooth edges and stress concentrations. 
In contrast, we target broad linear property coverage and inverse design with smooth surfaces.
}

\vspace{-0.3cm}
\subsection{Homogenization}
For shell homogenization, 
Schumacher et al.~\shortcite{schumacher2018mechanical} introduced a macroscale equivalent thin-shell model to derive parameters for mesoscale thin shells, 
while Yuan et al.~\shortcite{yuan2024volumetric} proposed a novel volumetric homogenization method specifically targeting knitwear. 
Additionally, Chen and Otaduy presented a high-order RBF interpolation technique in polar coordinates to address thin-shell homogenization~\cite{chan2024polar}.
Although our work also focuses on thin shells, 
we regard our building blocks as three-dimensional unit cells rather than patches, 
distinguishing our approach from prior studies. 
\change{Recently, data-driven models have emerged for efficient material prediction and design~\cite{Peng_2022,keshav2025spectral}. FFT-based homogenization methods have also been explored for cellular material design~\cite{Chen_2022}.}

We use a Finite-Element-based homogenization scheme~\cite{geers2010multi, panetta2015elastic, zhang2023optimized}, 
whereby linear elasticity is solved on a representative volume element (RVE).
We extend \cite{zhang2023optimized} to more efficiently support shells.

\vspace{-0.2cm}
\section{Design Space}
\label{sec:design-space}


\subsection{Surface Parameterization}


A shellular structure is often represented by a Signed Distance Field (SDF).
It is an implicit function in which the value zero identifies the surface.
Our objective is to parameterize the SDF in a low-dimensional space.
Inspired by electrostatics, 
we observe that an electric charge generates a potential field that depends on distance. 
Moreover, when there is an equal number of positive and negative charges, 
the resulting potential field closely resembles a SDF (see the inset).
In such a configuration, 
the surface appears at the location where the electric potential is zero. 
As a point moves farther from this zero-potential surface, 
the magnitude of the potential grows in a manner analogous to the SDF, \change{though it does not in general satisfy the Eikonal property.}

\noindent
\begin{minipage}[t]{0.45\columnwidth}
\setlength{\parindent}{1em}

Based on this observation, instead of defining a surface by an SDF, 
we represent the shellular structure through an electric potential field governed by a finite number of electric charges.
The resulting surface
\end{minipage}
\hfill
\begin{minipage}[t]{0.55\columnwidth}
\centering
\vspace{0pt}
\includegraphics[width=\linewidth]{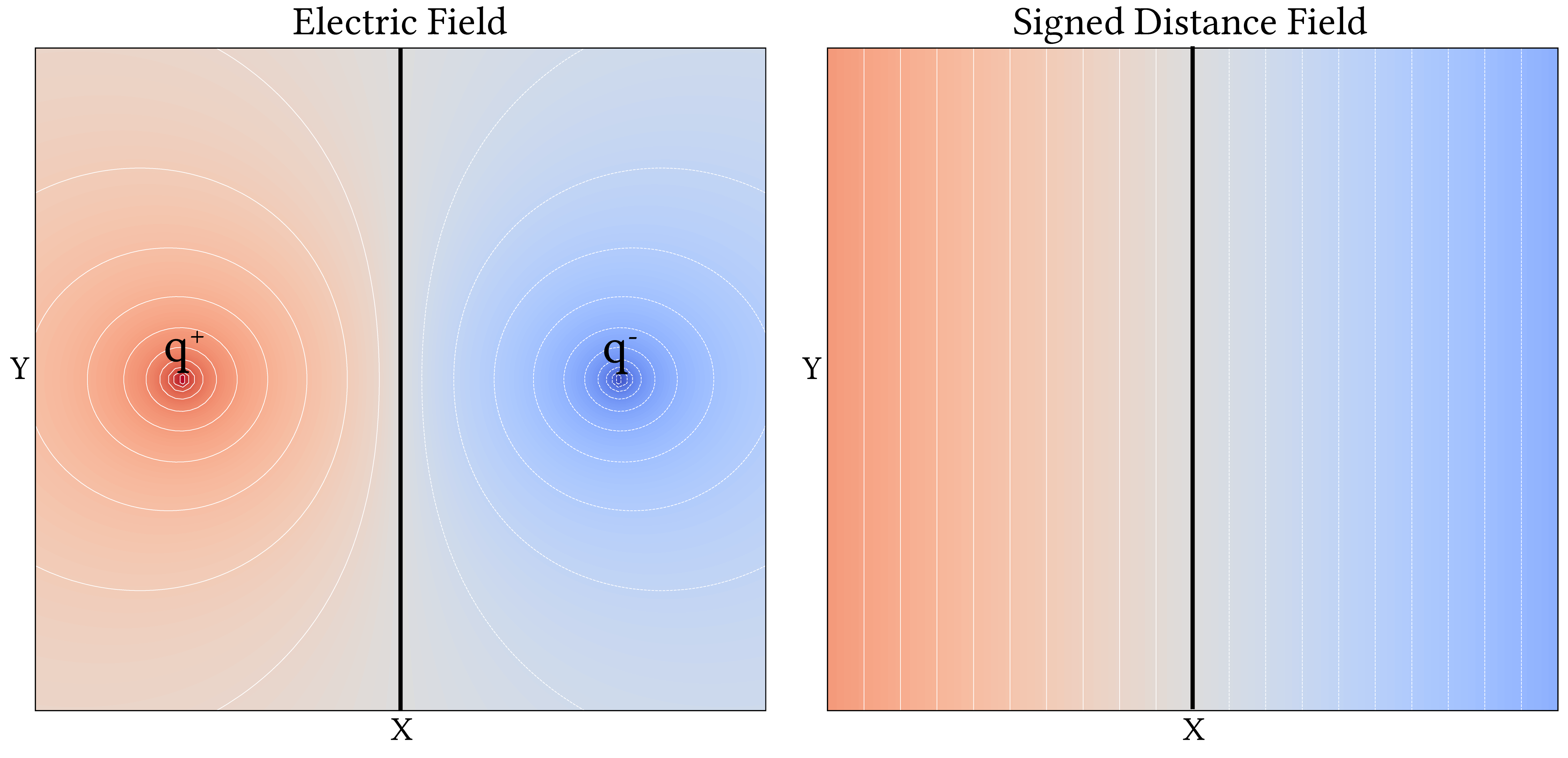}
\end{minipage}
is an equipotential surface. 
Hence, the design space for the shellular structure is characterized only by the positions and values of these charges, 
leading to a substantial reduction in the dimensionality of the control parameters.
The implicit function is written as
\begin{align}
F(x,y,z) = \iiint \frac{\rho(\xi,\eta,\zeta)\,d\xi\,d\eta\,d\zeta}{\sqrt{(x-\xi)^2 + (y-\eta)^2 + (z-\zeta)^2}},
\end{align}
where
\begin{align}
    \rho(x,y,z) = \sum_{j=1}^{N} q_j \,\delta\bigl(x-x_j,\,y-y_j,\,z-z_j\bigr)
\end{align}
is the charge distribution in 3D space, 
$\delta(\cdot)$ is the Dirac delta function, 
$(x_j,y_j,z_j)$ is the position of the $j$-th charge, and $q_j$ is its value.
For a shellular metamaterial, 
we impose periodic boundary conditions on the unit cell. 
Consequently, the electric charge distribution becomes a periodic function:
\begin{equation}
\rho(x,y,z) = \sum_{j=1}^{N} q_j \,\sum_{\substack{m,n,p=-\infty \\ m,n,p \in \mathbb{Z}}}^{\infty} \delta\bigl(x-m-x_j,\,y-n-y_j,\,z-p-z_j\bigr).
\end{equation}
Barnes~\shortcite{barnes1990microstructure} showed that, 
under these conditions in a unit cubic cell with period equal to 1, 
there is a closed-form expression for the resulting implicit function derived via a Fourier transform:
\begin{align}
&F(x,y,z) = \frac{8}{\pi}\sum_{j=1}^{N} q_j \sum_{\substack{h,k,l=0 \\ h,k,l \in \mathbb{Z}}}^{\infty} w_{hkl}B_{hkl}\bigl(x,y,z,x_j,y_j,z_j\bigr),\label{eq:inf-series}\\
B_{hkl}&\bigl(x,y,z,x_j,y_j,z_j\bigr) =\notag \\
&\frac{\cos(2\pi h (x-x_j))\cos(2\pi k (y-y_j))\cos(2\pi l (z-z_j))}{h^2 + k^2 + l^2},
\end{align}
where $w_{hkl} = 1/2$ if exactly one of $(h,k,l)$ is zero and $w_{hkl} = 1/4$ if two of the indices are zero. 



\change{Since the above series is infinite, it cannot be directly used to define a finite-dimensional design space.
We therefore truncate \(F\) to a finite series with \(2N\) charges and introduce weights
\(\alpha_{hkl}\) for \(h,k,l \le K\), with \(B_{000}(\cdot)=0\).
Each charge is assigned a sign \(q_j \in \{+1,-1\}\), with an equal number of positive and negative charges;
specifically, \(q_j=+1\) for \(j \le N\) and \(q_j=-1\) for \(j > N\).
This yields
\begin{equation}
\label{eq:charge-field}
F(x,y,z)
=
\sum_{h,k,l=0}^{K}
\alpha_{hkl}\, w_{hkl}
\sum_{j=1}^{2N}
q_j\, B_{hkl}(x,y,z,x_j,y_j,z_j).
\end{equation}}
In this formulation, the charge locations and the spectral weights $\{\alpha_{hkl}\}$ jointly determine the geometry. \change{Unless otherwise stated, all structures in this paper are defined by the zero level set $F(x,y,z)=0$.}

\noindent
\begin{minipage}[t]{0.51\columnwidth}
\setlength{\parindent}{1em}

The inset figure illustrates an example with $16$ charges for increasing truncation orders $K$, using
$\alpha_{hkl}=1$.
As $K$ increases, additional higher-frequency modes become available, 
allowing the surface to capture sharper geometric features and finer-scale details; 
at the same time, these higher-frequency components can introduce small-scale undulations.
By tuning $\alpha_{hkl}$, one can further refine the surface details.
As demonstrated in Sec.~\ref{sec:structure-gen}, 
even a relatively small number of charges and basis functions can produce a broad range of surfaces.
Moreover, because the dimension of this design 

\end{minipage}
\hfill
\begin{minipage}[t]{0.48\columnwidth}
\centering
\vspace{0pt}
\includegraphics[width=\columnwidth]{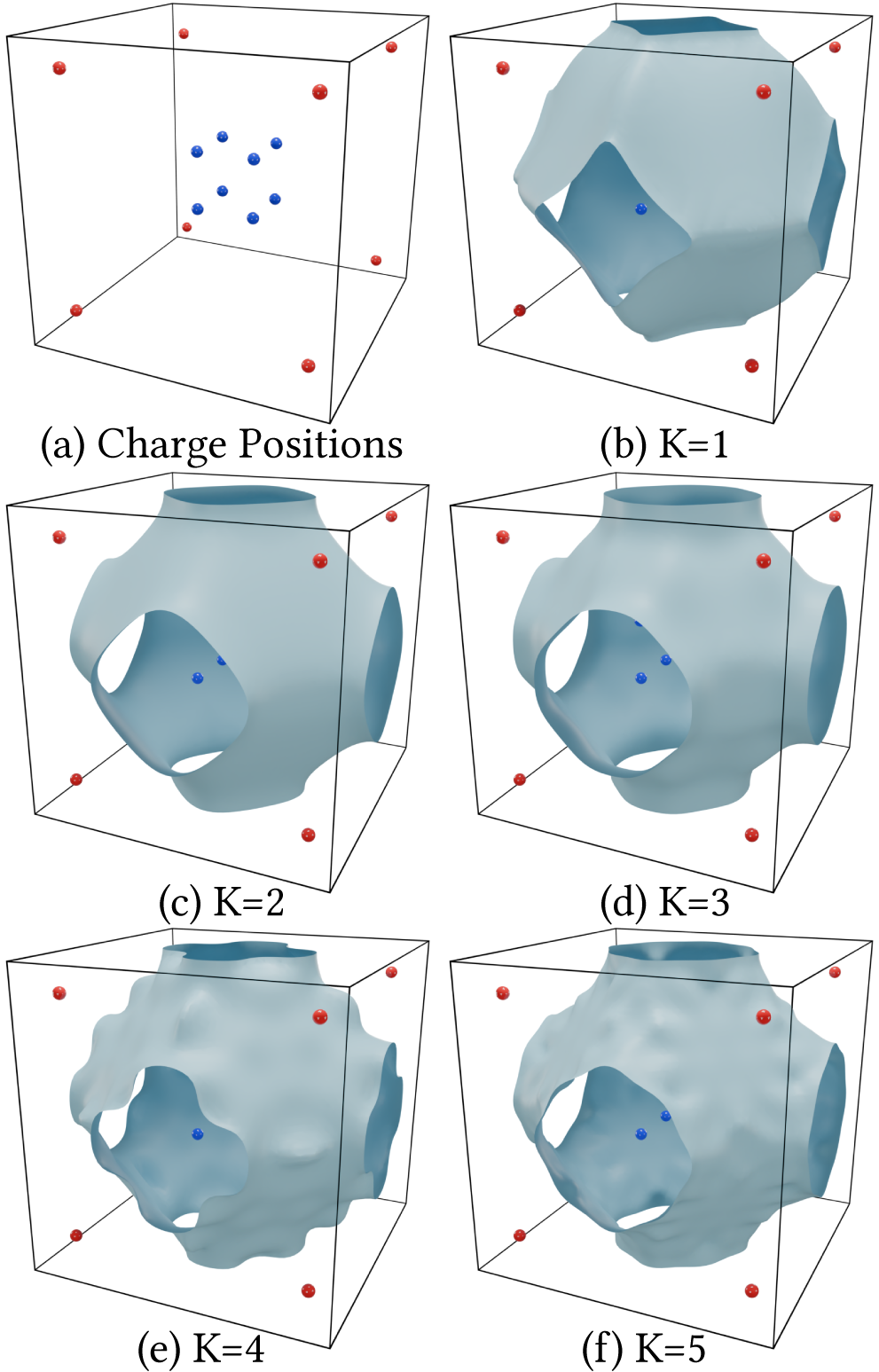}
\end{minipage}
space is low, 
the design process can be carried out efficiently using various optimization methods,
including first-order and gradient-free approaches.
Finally, this construction is closely connected to classical TPMS nodal approximations: 
Barnes~\shortcite{barnes1990microstructure} showed that truncating the periodic charge-series expansion (Eq.~\ref{eq:inf-series}) yields closed-form implicit functions that approximate several widely used TPMS families, 
suggesting that our parameterization naturally subsumes TPMS 
while generalizing beyond them through charge positions and spectral weights.

\vspace{-0.2cm}
\subsection{Structure Symmetry}
Building upon the existing design space, 
one can choose to impose structural symmetry by defining a \emph{Fundamental Bounding Volume} (FBV),
as other methods did. 
Rather than specifying charges throughout the entire unit cell, 
charges are assigned only within this FBV and then mirrored to span the full unit cell. 
This approach not only ensures structural symmetry but also reduces the dimensionality of the design space.
For example, although the figure above presents an example with 16 charges,
these charges are mirrored throughout the unit cell from only two charges located in a cubic FBV, 
yielding a symmetric final structure. 
Such structural symmetry often results in specific material properties, 
for instance, orthotropic behavior. 
In the Supplementary Material, we prove that symmetric charge configurations necessarily produce symmetric structures.
Typical FBVs include a region covering one-eighth of the unit cell or a tetrahedral domain, as discussed in~\cite{panetta2015elastic}.
In this paper, we use both of them as the FBVs.

\section{Efficient Homogenization}
\label{sec:homo}

\subsection{Voxel-based Homogenization of Thickened Shells}
To evaluate the effective elastic tensor of a shellular microstructure, 
we \emph{thicken} the implicit surface and embed the resulting geometry 
as a heterogeneous solid in a regular voxel grid over the unit cell.
Let $p_e=(x_e,y_e,z_e)$ denote the center of voxel element $e$.
We assign each voxel a soft occupancy $\beta_e\in[0,1]$ by sampling the implicit field $F$ and applying a smooth indicator function:
\begin{equation}
    K_e=\beta_e K_0,\ \beta_e=h(d_e),\ d_e=\frac{|F(\mathbf{p}_e)|}{\|\nabla F(\mathbf{p}_e)\|},
    \label{eq:beta}
\end{equation}
where $K_e$ is the voxel element stiffness matrix, 
$K_0$ is the constant stiffness matrix of a fully solid voxel (shared by all voxels), 
and $d_e$ is a first-order approximation of the (unsigned) distance from $p_e$ to the zero level set $F=0$.
The function $h(\cdot)$ controls how the zero level set is converted into a finite-thickness shell band.


\noindent
\begin{minipage}[t]{0.6\columnwidth}
\setlength{\parindent}{1em}
We use a logistic-smoothed Heaviside: \change{$h(v)=1/(1+\exp\left(-\kappa\,(t-v)\right))$},
where $t$ is the prescribed half-thickness and $\kappa$ controls the transition width.
\change{As shown in the inset, 
this formulation yields $h \approx 1$ for $v \leq t$ (solid) and $h \approx 0$ for $v \geq t$ (void), 
with a smooth transition around the surface.}
This volumetric embedding 

\end{minipage}
\hfill
\begin{minipage}[t]{0.39\columnwidth}
\centering
\vspace{0pt}
\includegraphics[width=\linewidth]{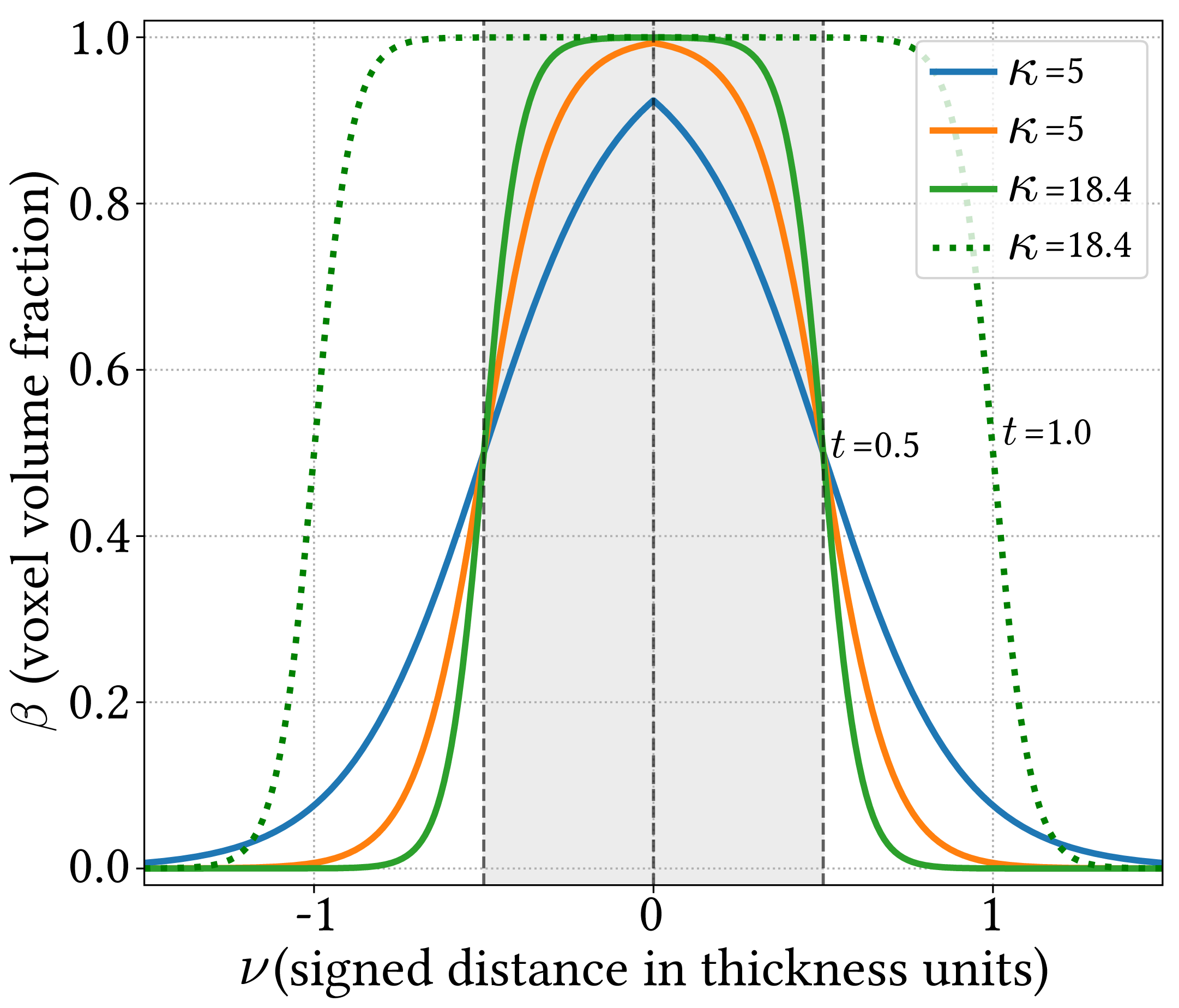}
\end{minipage}
 allows us to reuse standard periodic homogenization on structured grids 
while retaining a compact surface-based design parameterization.
The sharpness parameter $\kappa$ determines the width of the density transition; 
in our experiments we set $\kappa$ so that the transition spans approximately two voxels.
We choose $256^3$ resolution for shell homogenization accuracy based on 
a controlled study on an optimization task with known ground truth (Supplementary Materials).



\subsection{Sparse Multigrid Solver on Selective Voxel Grids}
A key computational challenge is repeatedly solving the voxelized periodic cell problems $\mathbf{K} \mathbf{u} = \mathbf{f}$
(six right-hand sides for linear elasticity homogenization).
Recent GPU voxel solvers (e.g.,~\cite{zhang2023optimized}) demonstrate that multigrid with efficient smoothers can provide high performance on dense voxel grids.
However, shellular structures are sparse in voxel space: only a thin band of voxels near the surface carries non-negligible stiffness.
Running a dense-grid solver therefore wastes compute on voxels that are effectively void and could cause accuracy issues.
To exploit this sparsity, we restrict computation to an \emph{active set} of voxels $\mathcal{A}^0=\{\,e\,|\,\beta_e > \varepsilon\,\},$
and extend the multigrid + colored Gauss-Seidel (GS) framework of Zhang et al.~\shortcite{zhang2023optimized} to operate on a
\emph{selective} (sparse) voxel hierarchy.

\begin{figure}
    \centering
    \includegraphics[width=\linewidth]{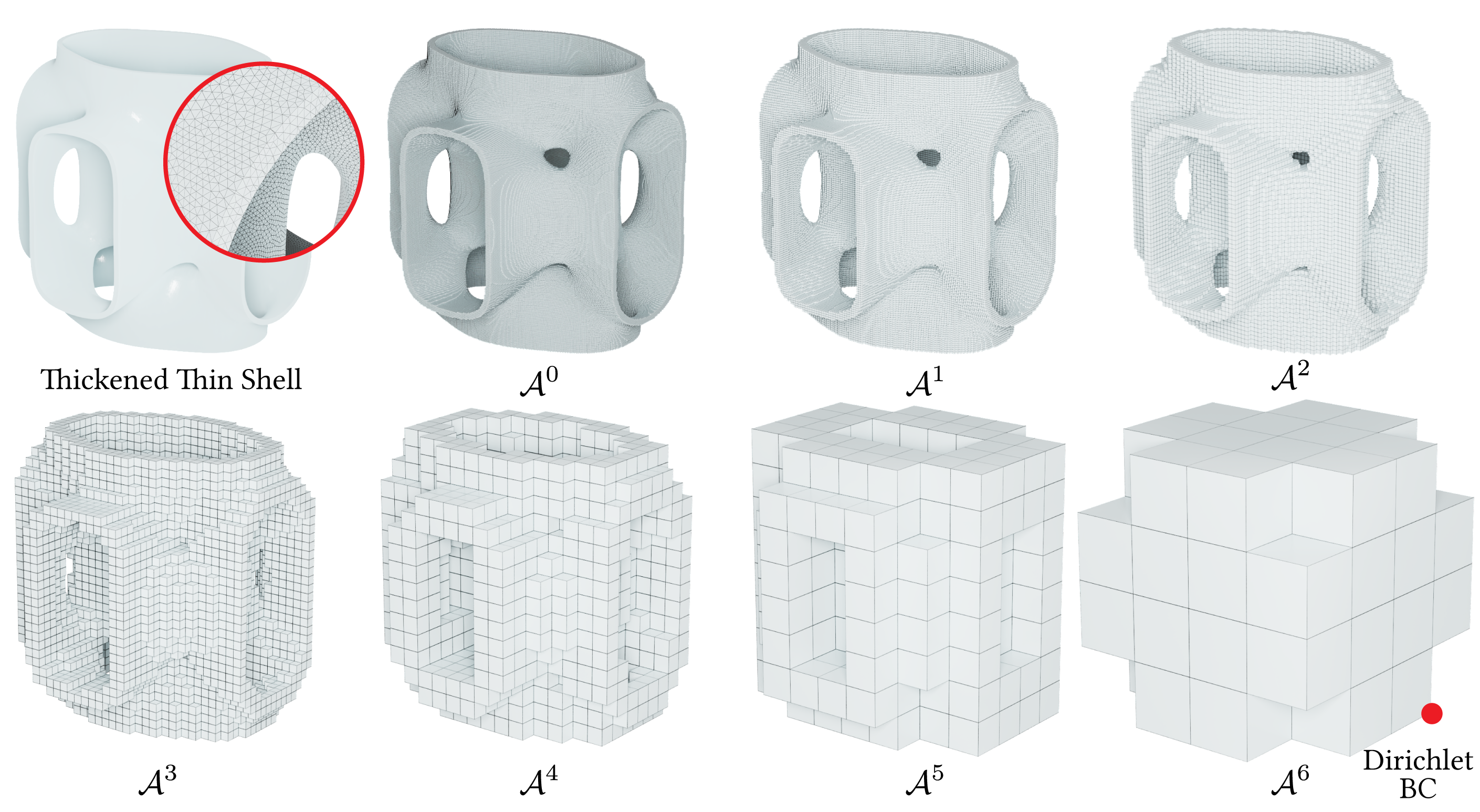}
    \vspace{-0.5cm}
    \caption{
    \textbf{Active voxels across multigrid levels.}
    We visualize the active-voxel hierarchy $\{\mathcal{A}^{\ell}\}$ used by our sparse multigrid solver.
    On the coarsest level (here, $\mathcal{A}^{6}$), 
    we impose a Dirichlet constraint on a single node to eliminate the nullspace 
    and solve the resulting system using a GPU dense direct solver.
    The inset (top-left) shows the conforming tetrahedral mesh, 
    which we use to compute the reference homogenized tensor $C$ for accuracy evaluation in  the commercial software.}
    \vspace{-0.5cm}
    \label{fig:voxels}
\end{figure}

\subsubsection{Solver Initialization}
Unlike dense-grid multigrid, 
a selective grid requires explicitly constructing the active hierarchy and the associated node sets.
We build a multilevel active voxel hierarchy $\{\mathcal{A}^{\ell}\}_{\ell=0}^{L}$ by standard $2\times2\times2$ coarsening:
a coarse voxel is marked active if any of its children are active (Figure~\ref{fig:voxels}).
From $\mathcal{A}^{\ell}$, we derive the corresponding set of \emph{active nodes} 
(vertices incident to at least one active voxel) 
and partition these nodes into color groups for multi-color GS relaxation on each level.
Each node is stored using a 32-bit integer $I$ that packs its integer grid coordinate $(i_x,i_y,i_z),$
$I=(i_z \ll 20)\,|\,(i_y \ll10)\,|\,i_x$, 
together with an 8-bit tag that encodes the node’s color-group ID and whether it is a boundary node.
For GPU efficiency, we sort active nodes within each color group by their encoded integer grid coordinates to improve memory coalescing.
Following Zhang et al.~\shortcite{zhang2023optimized}, 
we store all nodal vectors (displacements and residuals) in a structure-of-arrays (SoA) layout.
In addition, we store per-node stiffness blocks in the SoA format as well, 
which significantly improves memory throughput in the GS kernels.
Finally, to enforce periodic boundary conditions, 
we precompute the mapping between boundary nodes and their periodic
counterparts and allocate ghost-node tiles for fast memory access.

\subsubsection{V-cycle Implementation}
In practice, the dominant cost of GS relaxation is loading/storing stiffness data for each node.
We therefore adopt two strategies (building on~\cite{zhang2023optimized}):

\vspace{-0.2cm}
\paragraph{Stiffness re-usage.}
For the finest one level, we compute the local stiffness contributions on the fly to reduce GPU memory traffic.
For coarser levels, where the problem size is smaller, we compute once and store stiffness blocks in GPU memory,
avoiding the repetition of restriction computation and memory access.
The next time when the GS kernel is called for those levels, 
we directly load the stiffness blocks instead of computing it again.

\vspace{-0.2cm}
\paragraph{Batched solves for multiple right-hand sides.}
Homogenization requires solving six cell problems (corresponding to the independent macroscopic strain modes).
Instead of launching separate kernels for each right-hand side, we solve all six simultaneously within the same GS kernels.
This amortizes stiffness loads and per-node $3\times 3$ block factorizations across all right-hand sides.
We further track per-right-hand-side convergence (via residual norms) and skip updates for right-hand sides that have
already met a user-specified tolerance, further reducing the memory access.

Restriction and prolongation follow Zhang et al.~\shortcite{zhang2023optimized}: 
we restrict/prolong nodal displacements and residuals between levels per solving, 
which does not require additional stiffness access and remains effective on our selective grids.

\subsubsection{Accuracy and Performance}
Although stiffness access is the primary bottleneck, we store stiffness data in \texttt{fp32} for accuracy.
In our experiment, we found that the solver becomes more robust in high resolution voxels than using \texttt{fp16}. 
Compared to a dense-grid multigrid implementation on the full voxel mesh, 
our selective solver reduces runtime by avoiding computation in near-void regions while keeping the solution accurate.
At $256^3$ resolution with $t=0.02$, our selective solver achieves an average $\sim 8\times$ speedup over the dense baseline 
(about $0.35$\,s per homogenization in our setting) 
while producing effective properties within $6\%$ relative error compared to a reference analysis in \texttt{nTop} (Table~\ref{tab:time-cost})~\cite{ntop}.
This analysis uses a high-resolution periodic and conforming tetrahedral mesh of the thickened shell surface (Figure~\ref{fig:voxels}).
Notably, the dense-grid baseline requires assigning a nonzero stiffness floor to many near-void voxels, 
which can introduce additional stiffness bias; 
by explicitly culling voxels with $\beta_e\le \varepsilon$ (we use $\varepsilon=10^{-3}$), 
the selective solver reduces this effect.
Overall, our fast GPU homogenization enables interactive evaluation and 
makes inverse design over our compact parameter space practical; 
we demonstrate these applications in Sec.~\ref{sec:time-cost}.

\subsection{Inverse Design via Homogenization}
\label{sec:opt}
We formulate inverse design as an optimization over the charge locations $\{p_m\}_{m=1}^{2N}$, weights
$\{\alpha_{hkl}\}_{h,k,l=0}^{K}$, and the shell thickness $t$:
\begin{align}
\argmin_{\{p_m\},\{\alpha_{hkl}\},t}\  & \mathcal{E}\!\left(C^H\right),
\label{eq:inv-design}\\
\text{s.t.}\quad
& C^{H}_{ij} = \frac{1}{|Y|}\sum_{e\in Y} \bigl(\bm{\chi}^i+\mathbf{u}^{(i)}_e\bigr)^{\!T}\, K_e \,\bigl(\bm{\chi}^j+\mathbf{u}^{(j)}_e\bigr), \notag\\
& \mathbf{K}\,\mathbf{u}^{(k)}=\mathbf{f}^{(k)}, \quad i,j,k\in\{1,\dots,6\}, \notag\\
& K_e =\beta_e K_0,\ \beta_e=h\!\left(d_e(F(p_e; \{p_m\},\{\alpha_{hkl}\},t))\right),\ e\in Y. \notag
\end{align}
Here $Y$ denotes the voxel elements in the unit cell, 
$|Y|$ is the unit-cell volume, 
and $i,j$ index the six independent macroscopic strain modes (Voigt notation).
For each load case $k$, $\bm{\chi}^{k}$ is the ``test'' displacement and
$\mathbf{u}^{(k)}$ is obtained by solving the periodic linear system $\mathbf{K}\mathbf{u}^{(k)}=\mathbf{f}^{(k)}$,
assembled from element matrices $K_e$.
The objective $\mathcal{E}(\cdot)$ is user-specified and depends only on the homogenized tensor $C^H$.

\vspace{-0.2cm}
\paragraph{Differentiation.}
The volumetric embedding (Eq.~\ref{eq:beta}) makes the pipeline differentiable with respect to the design parameters
$\theta\in\{p_m,\alpha_{hkl},t\}$ through the voxel occupancies $\beta_e$.
Using the chain rule,
\begin{equation}
\frac{\partial \mathcal{E}}{\partial \theta}
\;=\;
\sum_{e\in Y}
\left(\frac{\partial \mathcal{E}}{\partial C^H} : \frac{\partial C^H}{\partial \beta_e}\right)
\frac{\partial \beta_e}{\partial \theta}.
\label{eq:chain}
\end{equation}
Because $\mathbf{u}^{(k)}$ satisfies equilibrium for each load case, 
$\partial \mathcal{E}/\partial \beta_e$ can be computed using standard methods as in topology optimization.
The remaining term $\partial \beta_e/\partial \theta$ is obtained analytically from $h(\cdot)$ and the implicit field
$F(\cdot)$  (see Supplemental Material for derivations).

\begin{figure}[t]
  \centering
    \includegraphics[width=\linewidth]{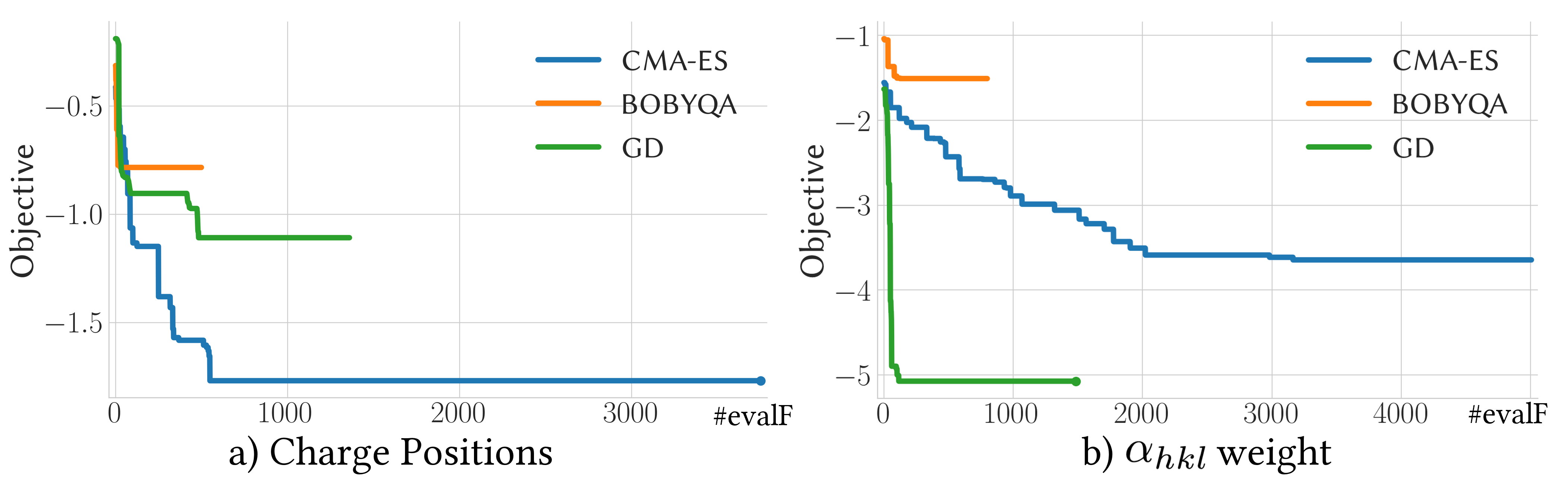}
    \vspace{-0.8cm}
  \caption{
  \textbf{Convergence comparison of optimization methods for different design variables.}
  Left: charge position optimization, where CMA-ES achieves lower objective values than gradient descent and BOBYQA.
  Right: $\alpha_{hkl}$ optimization, where gradient descent converges faster and outperforms gradient-free methods.
  }
  \vspace{-0.5cm}
  \label{fig:optimizer-comparison}
\end{figure}

\vspace{-0.2cm}
\paragraph{Choice of optimizers.}
Although gradients are available for all parameters, we found that optimizing charge positions $p_m$ is highly nonconvex
and can be effectively \emph{noisy} due to voxelization and the discrete active-set selection in the solver.
As shown in Figure~\ref{fig:optimizer-comparison}, first-order methods and local gradient-free optimizers (e.g., BOBYQA~\cite{powell2009bobyqa}) often stagnate early, 
whereas CMA-ES consistently discovers lower-energy designs~\cite{arnold2010active}.
We therefore optimize $\{p_m\}$ using CMA-ES.
In contrast, the weights $\{\alpha_{hkl}\}$ induce a smoother optimization landscape;
under the same objectives, gradient descent converges faster and to better values than gradient-free alternatives (Figure~\ref{fig:optimizer-comparison}).
Accordingly, we optimize $\{\alpha_{hkl}\}$ using a first-order method.

\vspace{-0.2cm}
\paragraph{Joint optimization with a volume constraint.}
In many tasks, users specify a maximum volume fraction.
We enforce this constraint by adjusting the shell half-thickness $t$ (via bisection) so that the voxelized structure meets
the target volume fraction, while optimizing $\{p_m\}$ and $\{\alpha_{hkl}\}$ for the objective.
Our full pipeline alternates three stages:
(i) update $t$ to satisfy the volume constraint,
(ii) optimize $\{\alpha_{hkl}\}$ for a fixed number of gradient steps with $\{p_m\}$ fixed,
and (iii) optimize $\{p_m\}$ using CMA-ES with $\{\alpha_{hkl}\}$ fixed.
We apply this alternating scheme to the design tasks in Sec.~\ref{sec:opt-result} and 
find it robust across a wide range of objectives.

\begin{figure*}
    \centering
    \includegraphics[width=\linewidth]{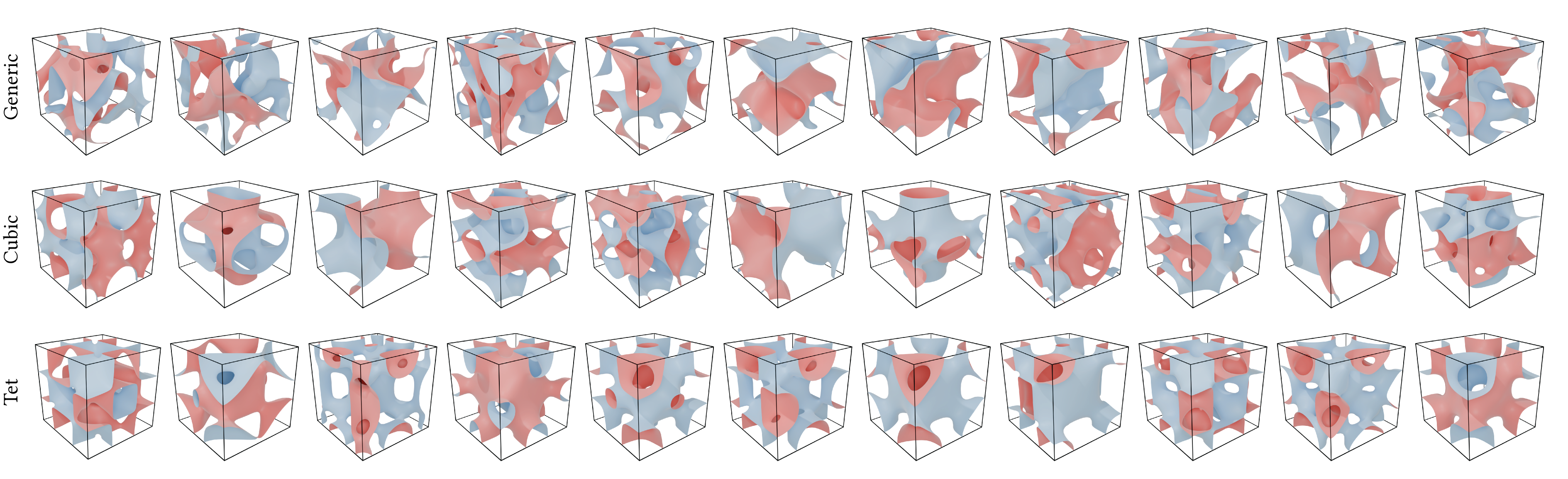}
    \caption{\textbf{Our design space can cover a large set of structures with diverse geometry.} Here we show three types of structures,
    structures constructed without symmetry, structures built from cubic FBV, and structures built from tetrahedron FBV.}
    \label{fig:random-surface}
\end{figure*}

\vspace{-0.2cm}
\section{Results}

\noindent
\begin{minipage}[t]{0.6\columnwidth}
\setlength{\parindent}{1em}
We comprehensively evaluate our method in terms of performance, 
structural diversity, structural property, and practical applications.
All experiments are conducted on a GPU server equipped with an NVIDIA H200~GPU, 
running Ubuntu~24.04 with CUDA~12.9.
The homogenization framework is fully implemented in C++ with CUDA. 
Although we vary the number of charges $2N$ according to particular applications, 
$K$ is fixed to 3 for all the examples presented.
\end{minipage}
\hfill
\begin{minipage}[t]{0.38\columnwidth}
\centering
\vspace{-0.5cm}
\setlength{\tabcolsep}{4pt}
\resizebox{\columnwidth}{!}{
\begin{tabular}{cc|c}
\hline
$r$ & $2N$ & $t_{\mathrm{field}}$ (ms) \\
\hline
64  & 32  & 0.7   \\
128 & 32  & 5.5  \\
256 & 32  & 43.0 \\
256 & 64  & 85.8 \\
256 & 128 & 171.7 \\
256 & 256 & 347.2 \\
\hline
\end{tabular}}
\captionof{table}{\textbf{Field computation cost.} $r$ denotes the voxel resolution, $2N$ the number of charges
after symmetry expansion, and $t_{\mathrm{field}}$ the time to compute the field.}
\label{tab:time-cost-field}
\end{minipage}

\vspace{-0.4cm}
\subsection{Computational Cost}
\label{sec:time-cost}

\begin{table}[!t]
\centering
 \resizebox{\columnwidth}{!}{
\begin{tabular}{ccc|cccc|cc|c|cc|c}
\hline
$r$ & $t$ &$n_e$ & $t_{\mathrm{a}}$ & $t_{\mathrm{rhs}}$ & $t_{\mathrm{solve}}$  & $t_{\mathrm{C}}$ & $t_{\mathrm{all}}$ & $t_{\mathrm{Zhang}}$ & $n_{\mathrm{it}}$ & $\epsilon$ & $\epsilon_{\mathrm{Zhang}}$ & $\mathrm{Mem}$\\
\hline
64  & 0.02 & 32K  & 8.6  & 0.1 & 117.6   & 0.1 & 126.4   & 377.0    & 5   & \change{6.6\%} & \change{13.8\%} & 1.1 \\
128 & 0.02 & 212K & 10.5 & 0.1 & 171.9   & 0.4 & 182.9   & 696.1    & 5   & \change{4.8\%} & \change{8.6\%} & 2.6 \\
256 & 0.02 & 1.5M & 16.7 & 0.1 & 336.5   & 2.4 & 355.7   & 2,919.4  & 5   & \change{4.1\%} & \change{5.4\%} & 13.5 \\
256 & 0.02 & 1.5M & 16.5 & 0.1 & 1,957.2 & 2.4 & 1,976.2 & 33,209.4 & 29  & \change{2.0\%} & \change{4.7\%} & 13.5 \\
256 & 0.04 & 2.8M & 18.0 & 0.2 & 376.9   & 4.5 & 399.6   & 3,091.3  & 5   & \change{1.5\%} & \change{5.7\%} & 13.5 \\ 
512 & 0.02 & 11.4M& 41.3 & 0.2 & 2,807.0 & 17.7& 2,866.2 & 37,129.6 & 15  & \change{1.4\%} & \change{7.5\%} & 100.4\\ 
\hline
\end{tabular}
}
\caption{\textbf{Time cost analysis of our method.} 
Column $r$ denotes the grid resolution, 
$t$ denotes the shell thickness, 
and $n_e$ denotes the number of active voxels.
Timing is broken down as follows: $t_{\mathrm{a}}$ (solver initialization), 
$t_{\mathrm{rhs}}$ (forming six right-hand sides),
$t_{\mathrm{solve}}$ (solve six displacements), 
$t_{\mathrm{C}}$ (computing the homogenized tensor $C$), 
and $t_{\mathrm{all}}$ (total runtime).
$n_{\mathrm{it}}$ is the number of V-cycles.
$\epsilon$ is the \change{relative $L^2$ norm error} with respect to the commercial reference solution, 
and $\mathrm{Mem}$ is the peak GPU memory usage in GB.
For comparison, $t_{\mathrm{zhang}}$ reports the runtime of the open-source solver of Zhang et al., 
and $\epsilon_{\mathrm{Zhang}}$ its relative error under the same setting.
All times are reported in milliseconds.
For the fourth-row experiment using Zhang et al. \shortcite{zhang2023optimized}’s method, 
we tightened the linear-solver tolerance by two orders of magnitude (i.e., set it to $100\times$ smaller than the default) to improve accuracy. 
\change{A more comprehensive comparison are provided in the Supplementary Material.}}
\vspace{-1cm}
\label{tab:time-cost}
\end{table}

\begin{figure}
    \centering
    \includegraphics[width=0.9\linewidth]{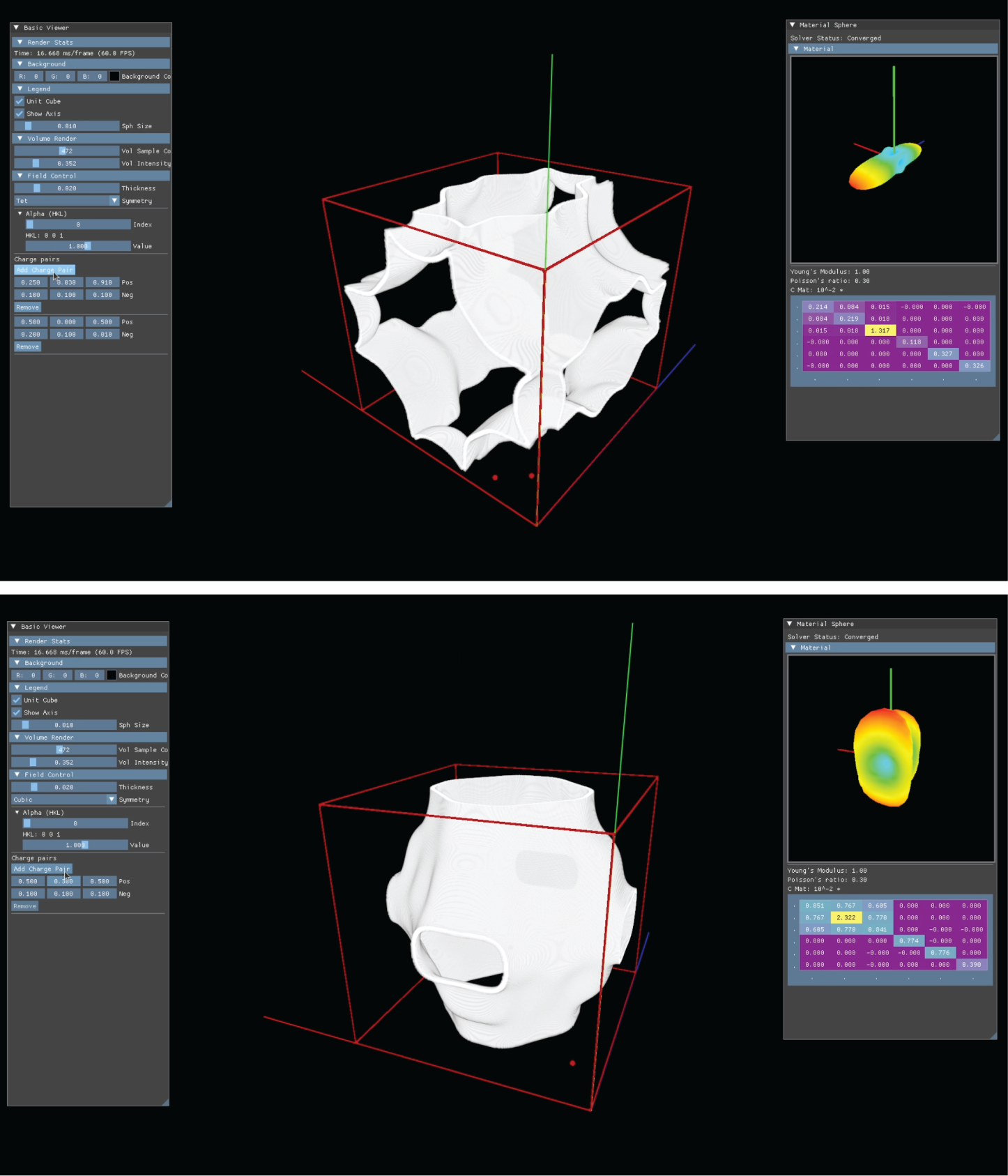}
    \caption{\textbf{Our graphical user interface (GUI).} 
    Users can interactively add, remove, or move charges to adjust the shell.
    Then our GPU-based multi-grid solver immediately computes the homogenized elastic tensor
    for user reference.}
    \vspace{-0.5cm}
    \label{fig:ui}
\end{figure}

Table~\ref{tab:time-cost} reports a breakdown of our homogenization runtime across voxel resolutions and shell thicknesses.
The time cost is measured on a dataset with 86 shells.
At $256^3$ resolution with thickness $t\in\{0.02,0.04\}$, 
our GPU implementation achieves near real-time evaluation (about $0.4\,\mathrm{s}$ per homogenization) 
while maintaining good agreement with the commercial reference (approximately \change{$4\%$} relative \change{$L^2$ norm} error).
As expected, increasing the number of V-cycles or using a higher voxel resolution can further improve accuracy, 
but at the cost of increased runtime.
Table~\ref{tab:time-cost-field} summarizes the cost of evaluating the implicit field $F$; 
even with a moderate number of charges, field evaluation takes only tens of milliseconds.
Together, this performance enables real-time, interactive exploration of shellular designs.
Figure~\ref{fig:ui} shows an example GUI in which the user edits the shell geometry and the corresponding effective
material properties are updated immediately, providing direct feedback during design.

\begin{figure*}
    \centering
    \includegraphics[width=\linewidth]{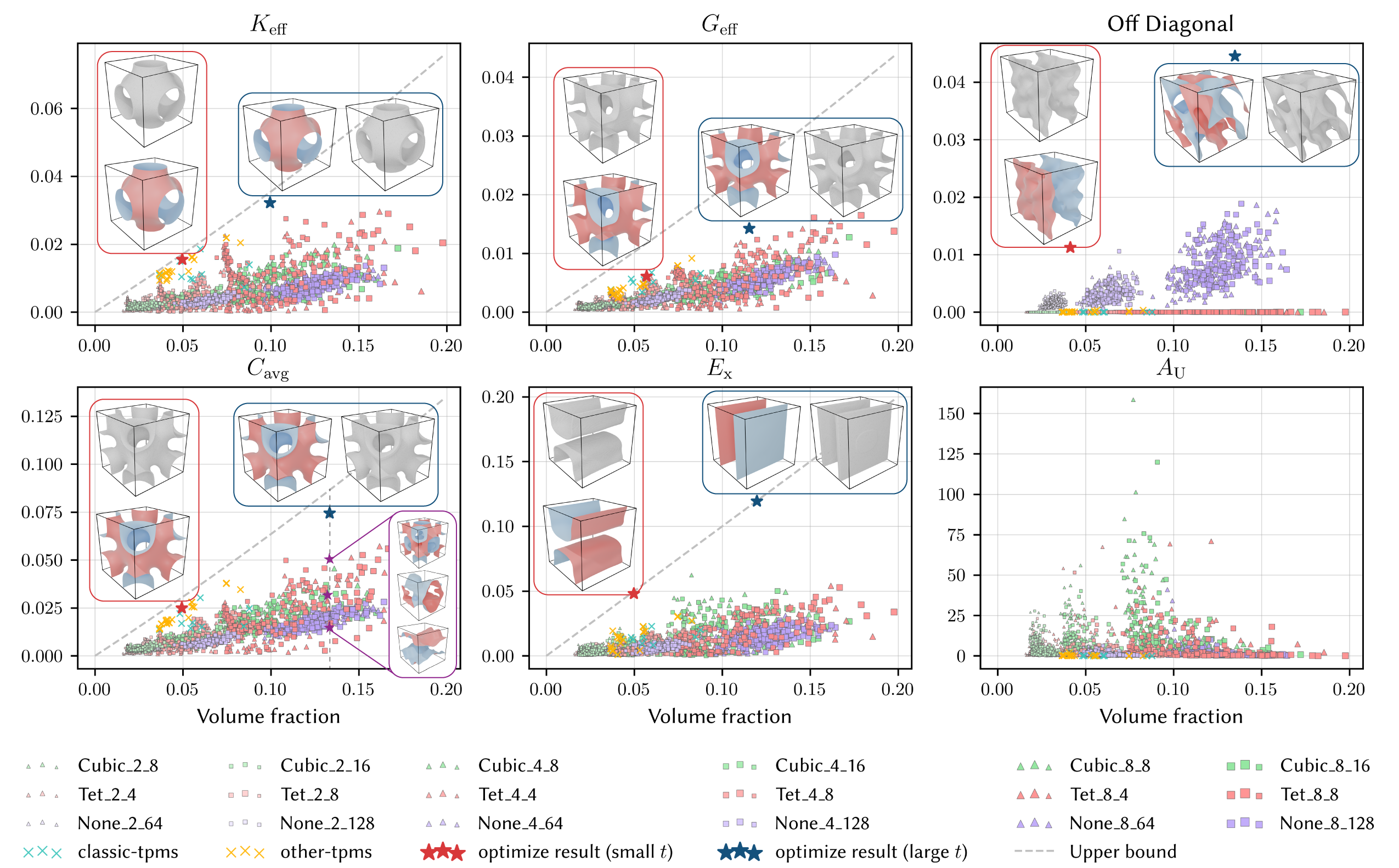} 
    \vspace{-0.5cm}
    \caption{
    \textbf{Our design space spans a wide range of effective material properties.}
    We evaluate each generated structure and report: 
    effective bulk modulus $K_{\mathrm{eff}}$, 
    effective shear modulus $G_{\mathrm{eff}}$,
    Young’s modulus along the $x$-axis $E_x$, 
    average normal stiffness $C_{\mathrm{avg}}$, 
    and the uniformity anisotropy index (UAI) $A_U$.
    ``Off Diagonal'' denotes the sum of absolute values of the entries in the upper-right $3\times3$ block of $C^H$ (Voigt form), indicating normal-shear coupling.
    Labels follow \texttt{sym\_$t$\_N}, denoting the symmetry mode, shell thickness in voxel-pitch units, 
    and the number of charges, respectively.
    Star markers indicate the final optimized designs for each objective (red/blue for the two thickness settings).
    Their corresponding shells and thickened structures are shown near the stars.
    For the $K_{\mathrm{eff}}$ objective, 
    the optimizer recovers a Schwarz~P-like shell, 
    consistent with prior observations that P-type TPMS achieve strong bulk-modulus performance~\cite{silvestre2007characterization}.
    For $C_\mathrm{avg}$, the dark purple stars indicate optimized designs targeting a set of approximately evenly spaced $C_\mathrm{avg}$ values while keeping the volume fraction fixed.
}
\vspace{-0.5cm}
    \label{fig:mat}
\end{figure*}

\vspace{-0.2cm}
\subsection{Random Structure Generation}
\label{sec:structure-gen}

We generate $18$ classes of structures by combining three symmetry modes
$\{\texttt{Cubic},\, \texttt{Tet},\, \texttt{None}\}$, 
three shell thicknesses $\{0.02,\,0.04,\,0.08\}$, 
and two charge-count settings.
For each class, we synthesize $150$ structures by randomly sampling charge positions, 
for a total of $2{,}700$ structures.
To quantify design diversity, we evaluate: 
Young's modulus along the $x$-axis $E_x$, 
the average normal stiffness $C_{\mathrm{avg}}=\tfrac{1}{3}(C_{11}+C_{22}+C_{33})$, 
the effective bulk modulus $K_{\mathrm{eff}}$, 
the effective shear modulus $G_{\mathrm{eff}}$, 
the universal anisotropy index (UAI)~\cite{ranganathan2008universal}, 
and the sum of absolute values of the off-block-diagonal entries of $C^H$ (Voigt form),
\change{which indicates normal-shear coupling}.
For $K_{\mathrm{eff}}$ and $G_{\mathrm{eff}}$, 
we use Hill’s approximation~\cite{hill1952elastic}.
We compute directional Young’s moduli from the compliance tensor as
$E_k = 1/((C^H)^{-1})_{kk}$.
Additional details are provided in the Supplementary Materials.

We compare our results against the theoretical Hashin--Shtrikman (HS) upper bounds for the $C_\mathrm{avg}, K_\mathrm{eff},G_\mathrm{eff}$~\cite{hashin1963variational}.
For $E_x$, we report the simple Voigt upper bound $vE$~\cite{voigt1910lehrbuch}, where $v$ is the volume fraction
and $E$ is the base material’s Young’s modulus, 
which is appropriate as a conservative upper bound even when the resulting microstructures are anisotropic.
A subset of the generated geometries is shown in Figure~\ref{fig:random-surface}, 
illustrating the shape diversity of our design space, 
while Figure~\ref{fig:mat} shows that the material properties span a broad range across all cases.

We also compare the expressiveness of our design space to (i) a truncated real Fourier basis and (ii) a Green’s-function
basis corresponding to periodic Laplace’s equation with point heat sources.
To match the number of design variables, we randomly sample $200$ complex-valued coefficient vectors for the Fourier basis (using a
$6\times 6\times 3$ set of coefficients) and generate heat-based fields with $64$ sources, and compare both to our
\texttt{None\_64} setting; all three parameterizations use approximately $192$ scalars.
Normalized by the corresponding theoretical upper bounds.

Our design space achieves a larger span and a higher maximum.
Against the Fourier basis, 
our span is $21.2\%$ vs.\ $6.9\%$ for $K_{\mathrm{eff}}$ (maximum $+4$ percentage-points (pp)) and $21.5\%$ vs.\ $3.1\%$ for $E_x$ (maximum $+16$ pp).
Against the heat (Green’s-function) basis, spans improve from $13.9\%\!\to\!21.2\%$ for $K_{\mathrm{eff}}$ and $15.1\%\!\to\!21.5\%$ for $E_x$, with maxima $+5$ pp for both.
\change{Against the star-shaped Voronoi method~\cite{numerow2025star}, 
generated from random lattice-based site arrangements with $N=8,12$, 
combining five metric-shape settings. 
Our method uses both \texttt{None\_8} and \texttt{None\_12} settings with five alpha configurations. 
Spans improve from $13.8\%\!\to\!25.9\%$ for $K_{\mathrm{eff}}$ (maximum $+13.4$ pp) 
and $15.7\%\!\to\!24.7\%$ for $E_x$ with a $+8.5$ pp maximum (see Supplementary Material for visual presentation).}

Finally, we compare a cubic-symmetry heat-based baseline with $8$ sources (placed in a cubic FBV) to our \texttt{Cubic\_8} setting.
Our method again achieves broader coverage and higher extremes: for $K_{\mathrm{eff}}$, the span increases from $36.6\%$ to
$41.9\%$ with a $+4$ pp maximum; for $E_x$, the span increases from $42.1\%$ to $75.1\%$ with a $+32$ pp maximum.


\subsection{Representing Existing Structures}

\begin{figure}
    \centering
    \includegraphics[width=1\linewidth]{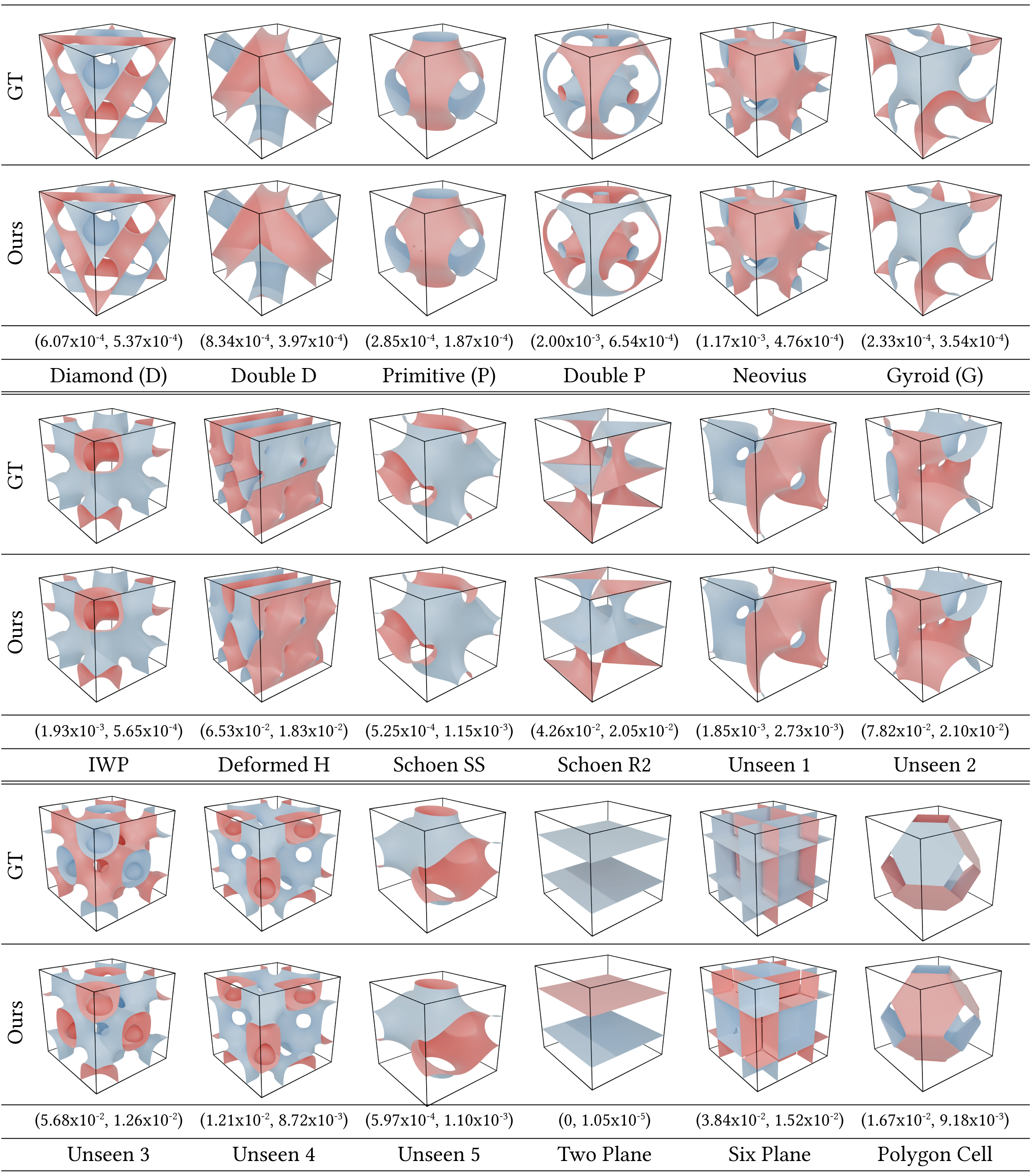}
    \vspace{-0.5cm}
    \caption{\textbf{Our design space encompasses existing structures.} 
    	In particular, we show 17 TPMS and three planar structures.
    	The small distance error above the name, represented as \change{relative $L^2$ elastic tensor error} and Hausdorff distance, 
    	underscores the accuracy of our representation.}
        \vspace{-0.5cm}
    \label{fig:existing-shells}
\end{figure}

Our approach can also capture TPMS and other shellular structures.
To achieve this, we first compute a set of TPMS configurations using~\cite{makatura2023procedural} or based on trigonometric functions. 
We then fit our representation to match these geometries by solving the following optimization problem:
\begin{align}
    \argmin_{p_m,\alpha_{hkl}}  \sum_i F^2\bigl(\bar{p}_i;p_m,\alpha_{hkl}\bigr) - \lambda \sum_i \tanh\Bigl(sF^2\bigl(\change{\bar{q}_i};p_m,\alpha_{hkl}\bigr)\Bigr),
\end{align}
where $\bar{p}_i$ denotes a randomly sampled point on the target surface, 
$\bar{q}_i$ is a randomly sampled point at a distance $d$ (we use $d\geq0.05$) from the target surface, 
$s$ is the inverse of absolute maximum of $F$ in each iteration (used to normalize $F$), 
and $\lambda$ is a hyperparameter. 
The first term ensures that our shellular representation passes through the points $\bar{p}_i$. 
The second term enforces separation from $\bar{q}_i$ by making $F$ nonzero.
We use the hyperbolic tangent function $\tanh$ because we do not want $F$ to grow arbitrarily large; 
it only needs to be sufficiently away from zero. 

Using this approach, 
we fit a considerable number of surfaces, 
including 50 TPMS and 3 planar surfaces, 
thereby demonstrating our method's ability to capture complex geometries.
A subset of results are presented in Figure~\ref{fig:existing-shells}.
To assess the accuracy of our reconstructions,
we compare the reconstructed surfaces with the reference geometries 
by computing both \change{the relative $L^2$ elastic tensor error} and the Hausdorff distance, 
as reported in Figure~\ref{fig:existing-shells}.
Our findings show that the resulting surfaces are closely matched with the ground truth, 
both visually and quantitatively. 
In addition, the properties of all fit surfaces are illustrated in Figure~\ref{fig:mat}.

\subsection{Optimal Structure Search}
\label{sec:opt-result}

We evaluate our inverse-design pipeline on six distinct optimization problems:
\begin{enumerate}
\item \textbf{Largest $E_{\mathrm{x}}$.} $\mathcal{E}(C^H) = -E_{x}/V$,  
where $V = \frac{1}{V_{\mathrm{all}}} \sum_{e} \beta_e$.
\item \textbf{Largest $C_{\mathrm{avg}}$.} $\mathcal{E}(C^H) = -\frac{1}{3}(\sum_{i=1}^{3} C^H_{ii}) / V$.
\item \textbf{Largest $K_{\mathrm{eff}}$.} $\mathcal{E}(C^H) = -K_{\mathrm{eff}}/V$.
\item \textbf{Largest $G_{\mathrm{eff}}$.} $\mathcal{E}(C^H) = -G_{\mathrm{eff}}/{V}$.
\item \textbf{Largest off-diag components.} $\mathcal{E}(C^H) = - \sum_{i=1}^{3} \sum_{j=4}^{6} \bigl| C^H_{ij} \bigr|$.
\item \textbf{User-specified force response.} $\mathcal{E}(C^H) = \bigl| C^H_{kk} - \bar{C}_k \bigr|$, where $\bar{C}_k$ is a user-defined value.
\end{enumerate}

The quantities $K_{\mathrm{eff}}$, $G_{\mathrm{eff}}$, and $C_{\mathrm{avg}}$ are computed as described in Sec.~\ref{sec:structure-gen}.
For $G_{\mathrm{eff}}$, 
we initialize from a randomly sampled design.
For all other objectives, we warm-start from the best design selected from the set of randomly sampled candidates
(Sec.~\ref{sec:structure-gen}) and then jointly optimize the coefficients $\alpha_{hkl}$ and charge positions using the
alternating pipeline in Sec.~\ref{sec:opt}, for both a thin-shell and a thick-shell setting.

The resulting optima are shown as red and blue stars in Figure~\ref{fig:mat}.
For $E_x$, the optimized designs reach $96.9\%$ and $99.8\%$ of the Voigt upper bound.
For $C_{\mathrm{avg}}$, we obtain two designs reaching $82.1\%$ and $86.1\%$ of the corresponding theoretical upper bound.
For $K_{\mathrm{eff}}$, the optimized values achieve $91.31\%$ and $91.40\%$ of the theoretical upper bound; 
while these are slightly below the best TPMS-like structure found in our design space (91.58\%), 
they remain the top performers at the corresponding volume fractions. 
For $G_{\mathrm{eff}}$, the optimized structures reach $52.3\%$ and $58.4\%$ of the upper bound.
These results are competitive with prior state-of-the-art performance in the low-density regime ($V \leq 10\%$) for
$C_{\mathrm{avg}}$, $K_{\mathrm{eff}}$, and $G_{\mathrm{eff}}$ (reported as $88.0\%$, $94\%$, and $52.8\%$, respectively)~\cite{ma2021elastically},
showing that a single design space can achieve strong performance across multiple objectives.
\change{
We also performed topology optimization at a small volume fraction of 0.1 for shear and bulk modulus~\cite{zhang2023optimized}. 
The optimal results are 38.2\% and 69.8\% of the theoretical upper bound, 
compared with 58.4\% and 91.5\% for our method.}
We additionally identify a design with large off-diagonal entries in $C^H$, indicating pronounced anisotropy.
In all cases, the optimized designs outperform random search, highlighting the effectiveness of our optimization pipeline.
\change{Our design space can produce nearly isotropic structures.
By setting our optimization objective to find the structure with the most isotropic elastic tensor, 
we are able to find a structure whose elasticity tensor differs from its closest isotropic tensor by only 0.22\%.}

Finally, we consider a user-specified directional force-response objective that emphasizes behavior along the $z$-direction.
Under this formulation, we obtain structures with targeted homogenized stiffness values
$C^H_{33} = 0.005$, $0.012$, and $0.035$.
We then tile these unit cells into larger blocks, as shown in Figure~\ref{fig:tile}.

\vspace{-0.2cm}
\subsection{Tiling and 3D Printing}

Our unit cells are periodic by construction and therefore tile seamlessly into larger volumes.
Moreover, heterogeneous blocks can be formed by assembling different unit cells within a single macroscopic part.
Because each structure is represented as an implicit function, 
we can blend neighboring designs by smoothly interpolating their implicit fields in a narrow band around block interfaces; 
the interpolation function and implementation details are provided in the Supplementary Materials.
Figure~\ref{fig:tile} shows an example that combines three distinct unit cells into a $5\times5\times5$ block, 
demonstrating large-scale tiling and cross-design composition.

\change{We also fabricate selected designs using a Digital Light Processing (DLP) 3D printer.}
We print three homogeneous samples, each consisting of a $5\times5\times5$ tiling of a single unit cell (one design per
block), as well as the heterogeneous block mentioned above.
All samples are printed successfully as $5\,\text{cm}\times5\,\text{cm}\times5\,\text{cm}$ cubes without support material,
with an average print time of approximately one hour per block.
We set the wall thickness to $0.015\,\text{cm}$, and the printed results are shown in Figure~\ref{fig:tile}.

\begin{figure}
	\centering
	\includegraphics[width=1\linewidth]{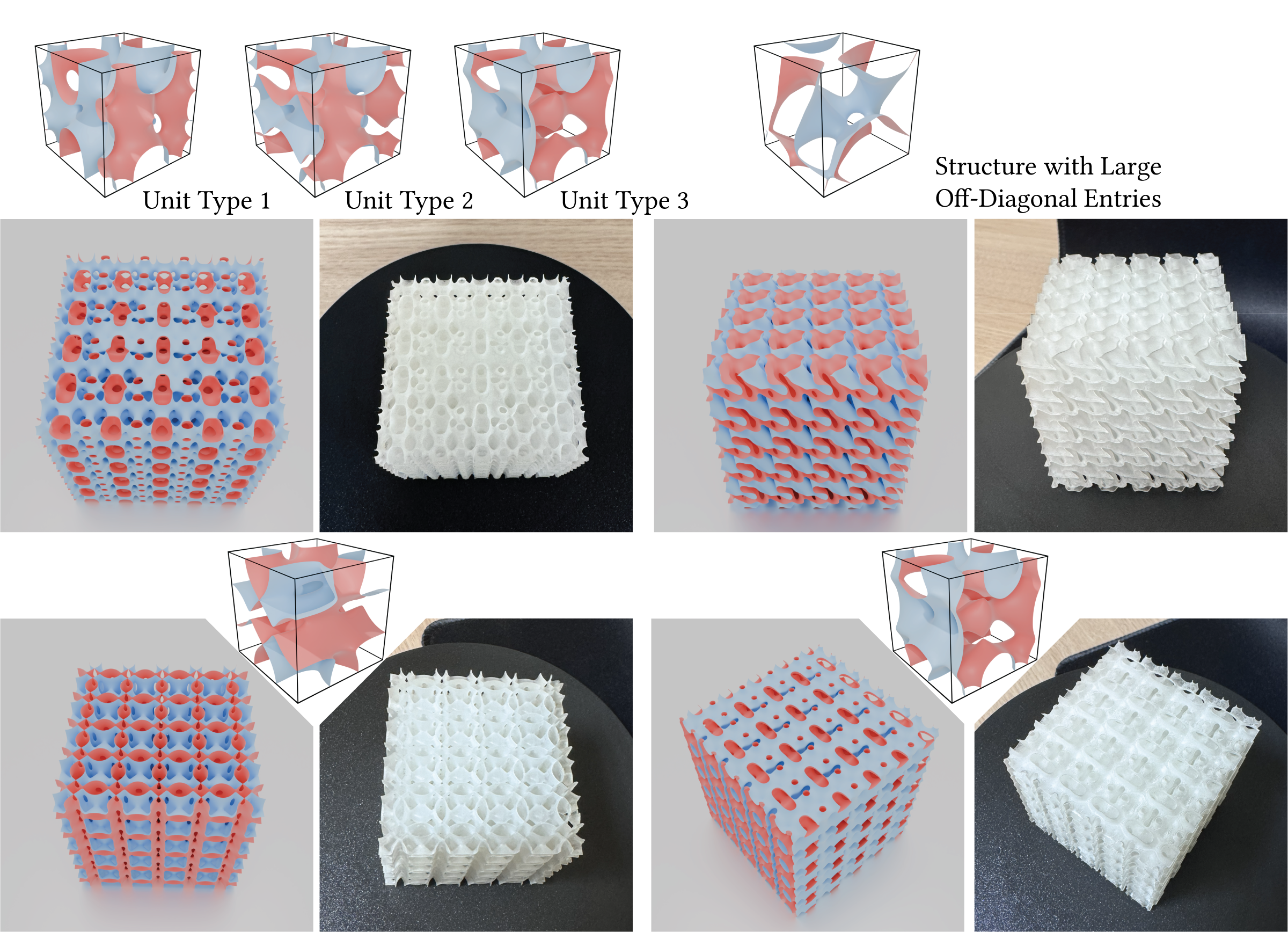}
	\caption{\textbf{Our structures are tileable and manufacturable.} 
		We present four distinct $5\times5\times5$ blocks assembled from our metamaterial design.
		The top-left block is composed of three different building blocks from 
        the inverse design given user-specified objectives.}
        \vspace{-0.5cm}
	\label{fig:tile}
\end{figure}


\vspace{-0.2cm}
\section{Conclusion}
We introduced a compact and versatile design space for shellular metamaterials, 
together with an efficient GPU pipeline for structure synthesis and homogenization-based property evaluation.
Our implementation enables near-real-time forward evaluation (typically under 0.5\,s per unit cell) and supports
inverse design for both property-driven objectives and user-specified directional force responses.
Extensive qualitative and quantitative experiments demonstrate that our representation spans a diverse family of
shellular geometries and achieves strong effective mechanical performance under multiple metrics.
We further validated manufacturability by 3D printing tiled blocks, highlighting the potential of our approach for
practical applications such as lightweight high-stiffness components and metamaterial infills.

Our method currently estimates effective properties via voxel-based embedding and homogenization, 
which introduces discretization error, especially for very thin shells.
A natural next step is to integrate shell- or surface-based simulation to improve accuracy 
while preserving interactive performance.
\change{Our formulation does not explicitly enforce connectivity constraints.
If needed, connectivity constraints could also be incorporated into the optimization, 
for example through penalties on isolated components.}
Beyond mechanics, our implicit, periodic representation may also be applicable to other physics-driven design tasks,
including thermal and fluid objectives.
Finally, systematically characterizing real-world performance -- including nonlinear effects, buckling, and material
imperfections -- remains an important direction for future work.

\bibliographystyle{ACM-Reference-Format}
\bibliography{citations}

\end{document}